\documentstyle[preprint,aps]{revtex}

\begin{document}
\draft

\title{Possible three-nucleon force effects in D--P scattering at low energies}

\author{ C.~R.~Brune$^1$, W.~H.~Geist$^1$,
  H.~J.~Karwowski$^1$, E.~J.~Ludwig$^1$, K.~D.~Veal$^1$,
  M.~H.~Wood$^1$, A.~Kievsky$^2$, S.~Rosati$^{2,3}$, and M.Viviani$^2$}
\address{ $^1$Department of Physics and Astronomy, University of North 
  Carolina at Chapel Hill, Chapel Hill, North Carolina 27599-3255
  and Triangle Universities Nuclear Laboratory, Durham, 
  North Carolina 27708 }
\address{ $^2$Istituto Nazionale di Fisica Nucleare, Piazza Torricelli 2,
  56100 Pisa, Italy }
\address{ $^3$Dipartimento di Fisica, Universita' di Pisa,
  Piazza Torricelli 2, 56100 Pisa, Italy }

\date{\today}

\maketitle

\abstract{We present measurements of the analyzing powers
$A_y$ and $iT_{11}$ for proton--deuteron scattering at
$E_{c.m.}=432$~keV.} Calculations using a realistic nucleon-nucleon
potential (Argonne V18) are found to underpredict
both analyzing powers by $\approx 40\%$.
The inclusion of the Urbana three-nucleon interaction
does not significantly modify the calculated analyzing powers.
Due to its short range,
it is difficult for this three-nucleon interaction to
affect $A_y$ and $iT_{11}$ at this low energy.
The origin of the discrepancy remains an open question.}

\narrowtext
\noindent{key words: N-d scattering, polarization observables}\\

\noindent{Corresponding author: Carl R. Brune} \\
\noindent{Triangle Universities Nuclear Laboratory,}\\
\noindent{Duke University, Box 90308, Durham, N.C., USA 27708-0308} \\
\noindent{tel: (919)660-2620, Fax: (919)660-2634,
  email: carlb@physics.unc.edu} \\

\newpage

Few-body systems provide a fundamental testing ground
for nuclear interactions.
Comparisons of measured three-nucleon scattering observables
to theoretical calculations allow stringent tests of
the underlying nucleon-nucleon (NN) and models of
the three-nucleon (3N) interactions.
Past studies have found that rigorous calculations
utilizing realistic NN potentials underpredict by 25-30\%
the measured analyzing power $A_y$ in n-d scattering at low energies --
a surprising discrepancy which has been dubbed the ``$A_y(\theta)$ puzzle''
(see Ref.~\cite{Wit94} and references therein).
Precise calculations for p-d scattering~\cite{Kie94,Kie95} below the
deuteron breakup threshold including the Coulomb potential rigorously
have recently become possible to perform.
These calculations find~\cite{Kie95,Kie96} that a
similar underprediction exists here for $A_y$ and also $iT_{11}$.
The agreement for other observables, including cross sections,
$T_{20}$, $T_{21}$, $T_{22}$, and n-d scattering lengths~\cite{Kie97}
is generally excellent.
The analyzing powers $A_y$ and $iT_{11}$ in N-d scattering
are known to be very sensitive to the
NN potential in the $^3P_j$~waves.
It has been suggested that the potential in these waves may not be known to
the necessary precision at low energies~\cite{Wit91,Tor97}.
The possibility that the underprediction is due to 3N force effects
has also been considered~\cite{Wit94,Kie95,Kie96}.
This paper investigates the roles of NN and 3N force effects on
$A_y$ and $iT_{11}$ for N-d scattering; we do not consider here
other possibly important effects such as relativistic corrections
or subnucleonic degrees of freedom.
We do note, however, that the Mott-Schwinger interaction, a long-ranged
electromagnetic effect, has been recently shown {\em not} to be
responsible for the discrepancies in $A_y$ and $iT_{11}$~\cite{Sto97}.

Measurements of both $A_y$ and $iT_{11}$ are useful, as these observables
depend on different combinations of phase shift and mixing parameters
(principally in $P$-waves), and they also have been shown to be sensitive
to different combinations of the $^3P_j$ NN interactions~\cite{Wit91}.
The majority of the data on these analyzing powers has been obtained
in the vicinity of the deuteron breakup threshold:
n-d $A_y$ data exist for $E_{c.m.}\ge 2$~MeV~\cite{Tor91,Mca94}
p-d $A_y$ data exist for $E_{c.m.}\ge 0.53$~MeV~\cite{Hut83,Knu93,Shi95},
and p-d $iT_{11}$ data exist for $E_{c.m.}\ge 1.7$~MeV~\cite{Knu93,Shi95}
(there are no n-d $iT_{11}$ data).
It is desirable to determine these observables at lower energies,
as the influence of higher partial waves is strongly reduced, and the
dominant $S$- and $P$-waves can be investigated with more confidence.
In addition, as we will show below, $P$-wave N-d scattering at low energies
is almost entirely determined by the asymptotic part of the
three-nucleon scattering wave function and the NN interaction.
Under these conditions the connection between the measured observables
and the underlying interactions is greatly simplified.
The calculated analyzing powers $A_y$ and $iT_{11}$
are mainly determined by the $j$-splitting of
the $P$-wave N-d phase shifts and the $\varepsilon_{3/2}^-$
and $\varepsilon_{1/2}^-$ mixing parameters.
Due to the angular momentum barrier at low energies, these
observables are very small and difficult to determine experimentally.
This paper reports measurements of $A_y$ and $iT_{11}$
for p-d elastic scattering at $E_{c.m.}=432$~keV.
These data are at the same energy as our previously-reported
$T_{20}$ and $T_{21}$ measurements~\cite{Kie97}.

These experimental results are compared to calculations
utilizing the Pair-Correlated Hyperspherical Harmonic
(PHH) basis~\cite{Kie93} to construct the scattering wave function,
and the Kohn variational principle to determine
the scattering matrix elements~\cite{Kie94}.
In addition, we present calculations using an ``optimized'' 
Born approximation~\cite{Kie96} for the peripheral partial waves.
The calculations have been done using the AV18 potential~\cite{AV18}
and with AV18 plus the 3N interaction
of Urbana (UR)~\cite{UR}.
It has been shown in Ref.~\cite{Wit94} that other high-quality NN
potentials such as Bonn or Nijmegen predict essentially the same
n-d $A_y$ just below the deuteron breakup threshold, so we
would not expect our conclusion to change if these potentials were used.

The measurements were performed using polarized proton and deuteron
beams from the atomic beam polarized ion source~\cite{Cle95} at
the Triangle Universities Nuclear Laboratory. The deuteron beams were
accelerated to $E_d=1.3$~MeV using the FN tandem accelerator, and
then directed into a 62-cm diameter scattering chamber.
Proton beams were accelerated into a 107-cm diameter scattering chamber
at $E_p=650$~keV using the minitandem accelerator~\cite{Bla93}
and chamber bias voltage~\cite{Lud97}.
Thin hydrogenated or deuterated carbon targets were utilized which consisted
of approximately $1\times 10^{18}$ and $1.5\times 10^{18}$ hydrogen isotope
and carbon atoms/cm$^2$, respectively.
The beams lose $\approx 10$~keV in these targets,
with an average energy of $E_{c.m.}=432\pm 1$~keV for both measurements.
The use of thin targets is very important at low energies for
minimizing energy loss and straggling effects.

The proton beam polarization was determined using the
${}^6{\rm Li}(\vec{p},{}^3{\rm He}){}^4{\rm He}$ reaction in
a polarimeter~\cite{Bru97} located at the rear of the scattering chamber.
The polarization was measured several times throughout the measurements
at an incident proton energy of 450~keV by lowering the
chamber bias voltage. The proton polarization was found to be constant within
$\pm 3\%$ throughout the measurements; the systematic error in the
proton polarization is estimated to be $\pm 4\%$.
Deuteron beam vector polarization was determined online via the
${}^{12}{\rm C}(\vec{d},p)$ reaction in a polarimeter
located behind the scattering chamber.
The effective $iT_{11}$ for this reaction at $E_d=1.3$~MeV
has been calibrated relative to the ${}^3{\rm He}(\vec{d},p)$
reaction in another polarimeter at $E_d=12$~MeV~\cite{Ton80}.
The absolute uncertainty in the deuteron beam polarization is
estimated to $\pm 3\%$.
For both beams the data were taken with the spin-quantization axis
perpendicular to the reaction plane, using two spin states 
with $p_{Z}\approx\pm 0.7$ for the proton beam; and
$p_{Z}\approx\pm 0.55$, $p_{ZZ}\approx 0$ for the deuteron beam.
The spin states were cycled approximately once every second, in order
to minimize the effects of slow changes in beam position, target
thickness, or amplifier gain.

Scattered deuterons and protons were detected in coincidence using
two pairs of silicon surface barrier detectors placed at symmetric
angles on either side of the incident beam.
The angles of the detectors were set
to observe either protons or deuterons in the
more forward detectors in coincidence with deuterons or protons
detected in the more backward detector on the opposite side of the beam.
Histograms of the time difference between the fast timing signals
from each coincident pair of detectors were stored for each spin state.
Dead-time corrections ($<3\%$) were determined by
sending test pulses to the detector preamplifiers with time delays
adjusted to give distinct peaks in the time spectra.
The time resolution for the coincident proton-deuteron peaks
was $\approx 10$~ns, with backgrounds $<3\%$.
The analyzing powers were determined from the counts in the coincident
peaks, after correction for background, dead time,
and the number of incident particles (determined by beam-current integration).
It should be noted that the coincidence technique is essential
for measuring the small analyzing powers $A_y$ and $iT_{11}$
with these targets, as the elimination of carbon elastic-scattering
events by the fast coincidence requirement allows proton-deuteron
scattering events to be counted at the high rate required to
achieve reasonable statistical accuracy.
The results for $A_y$ and $iT_{11}$ are shown in Fig.~\ref{fig:vector}.
The error bars include contributions from statistics and background
subtraction, but not the absolute beam polarization.

The theoretical method has been described previously~\cite{Kie94,Kie95};
it can be applied equally well to n-d as well as p-d scattering,
and realistic NN and 3N potentials can be used without difficulty.
In the present calculations, scattering waves with orbital angular
momentum up to $L=4$ have been taken into consideration.
At this energy, the differential cross section,
$A_y$, and $iT_{11}$ are determined almost entirely by
waves with $L\le 1$, while for $T_{20}$, $T_{21}$, and $T_{22}$ 
$L=2$ waves are also  important.
In particular, $A_y$ and $iT_{11}$ change by $<10^{-4}$ when
phases with $L>1$ are considered.
In Fig.~\ref{fig:vector} the data are compared to the calculations
using the AV18 potential and the AV18+UR potential.
The corresponding $P$-wave and $^4S_{1/2}$ phase-shift parameters
are given in Table~\ref{tab:phase}.

It is seen that both calculations underpredict the data by $\approx 40\%$.
The change in the calculated $A_y$ and $iT_{11}$
resulting from the inclusion of the 3N interaction
is too small {\em by an order of magnitude} to explain the discrepancy.

Relatively small changes in the N-d phase shift parameters can
have large effects on the corresponding analyzing powers.
In Ref.~\cite{Kie96} it was found that the discrepancies in
$A_y$ and $iT_{11}$ for $E_{c.m.}=1.67$ and 2~MeV
could be corrected by reducing the 
$^4P_{1/2}$ phase shift by 3.4\% and increasing the absolute value of
$\varepsilon_{3/2}^-$ mixing parameter by 12\%.
Using the AV18+UR results for the other phase-shift parameters,
we find that agreement with our $A_y$ and $iT_{11}$ results is optimized
if the $^4P_{1/2}$ and $\varepsilon_{3/2}^-$ parameters given in
Table~\ref{tab:phase} (column 2, in parenthesis)
are replaced by $5.22^\circ$ and $-1.02^\circ$,
i.e., the absolute values are reduced by 1.6\% and
increased by 15\%, respectively.
The results for $A_y$ and $iT_{11}$ using these parameters are
shown by the long-dashed curve in Fig.~\ref{fig:vector}.
It is important to note that these parameter changes affect the
cross section by $<0.15\%$, and the other analyzing powers by $<0.0012$.
In particular, the good agreement observed previously with
our $T_{20}$ and $T_{21}$ data at this energy~\cite{Kie97} is not disturbed.
While similar to the changes required at higher energies~\cite{Kie96},
there are significant differences in the fractional changes required.
We should point out however that there is no reason to expect
the percentage change required to be the same for different energies.

We have also performed calculations using an ``optimized''
Born approximation, i.e., the procedure in which
the second-order ${\cal R}$-matrix is 
estimated using the asymptotic part of the three-nucleon wave function
as described in Ref.~\cite{Kie96}. 
The results for the $^4S_{3/2}$ and $P$-wave phase-shift parameters are given
in Table~\ref{tab:phase} and compared to those obtained when the complete
wave function is considered (full solution). 
It is seen that the Born approximation results are
close to the full calculation for these partial waves.
In the case of the $^4S_{3/2}$ phase the Pauli principle prevents the
three particles from being close to each other, while for the $P$-waves the
centrifugal barrier is sufficiently high at these energies.
These findings indicate that these partial waves
are almost entirely determined by the asymptotic structure of the system.
We also show in Table~\ref{tab:phase} the results for the full solution
and Born approximation at $E_{c.m.}=2$~MeV, where it is seen
that the accuracy of the Born approximation for low partial waves
is reduced. On the other hand we observe that the influence of the 3N force
is small and of the same magnitude at both energies.

The small effect on $A_y$ and $iT_{11}$ from including
the UR 3N interaction is now clear.
This interaction, which is based on two--pion exchange and includes a
phenomenological repulsive short range term,
requires the three nucleons to be close together in order to produce
a significant effect.
The likelihood for this situation is diminished by the diffuse structure
of the deuteron which results from the small binding energy.
For low energy p-d scattering in $P$-waves (or higher $L$ values),
the probability of finding three nucleons in close proximity is
further reduced by the centrifugal and Coulomb barriers.
We thus draw the important conclusion that
3N interactions based on two-pion exchange cannot produce significant changes
in $A_y$ and $iT_{11}$ at low energies.

Other choices for the 3N potentials, such as the Tucson-Melbourne~\cite{TM}
or the Brazil~\cite{BR} models, give quite similar conclusions. 
Inclusion of other processes, such as $\pi-\rho$ or $\rho-\rho$ exchanges, 
involving heavier mesons and therefore
shorter ranges, are expected to give still smaller corrections~\cite{Wit94}.
These findings thus indicate that new types of 3N interactions should
be considered. One possibility is the inclusion of a spin-orbit 3N force
which could significantly affect the N-d $P$-waves~\cite{TNLS}.
One cannot exclude also the possibility that inadequacies in the NN
interaction are responsible for the discrepancy.

In summary, our measurements of $A_y$ and $iT_{11}$ at $E_{c.m.}=432$~keV
are significantly underpredicted by calculations utilizing the AV18
NN interaction. The inclusion of the UR 3N interaction does not significantly
change the theoretical calculations.
We have shown it is difficult to identify a 3N interaction
which could significantly
change these analyzing powers at low energies,
as they are mainly determined by long-ranged interactions.
It would be of great interest to extend these comparisons to
p-$^3{\rm He}$ and n-$^3{\rm H}$ scattering, where the 3N force
effects are expected to be larger, as the likelihood of finding three
nucleons close together is enhanced by the tighter binding
of $^3{\rm H}$ and ${}^3{\rm He}$.

\section*{Acknowledgments}

The authors would like to thank B.~M.~Fisher
for assistance in the data collection process.
One of the authors (A. K.) would like to thank Duke University and 
TUNL for hospitality and partial support
during his stay in Durham, where part of the present work was performed.
This work was supported in part by the U.S. Department of 
Energy, Office of High Energy and Nuclear Physics, under 
grant No. DE-FG02-97ER41041.

\begin{figure}
\caption{Experimental $A_y$ and $iT_{11}$ for p-d scattering
at $E_{c.m.}=432$~keV (circles),
along with theoretical calculations using the AV18 (solid line)
and the AV18+UR potentials (short-dashed line). The long-dashed
line results from modifying the $^4P_{1/2}$ and $\varepsilon_{3/2}^-$
phases as described in the text.}
\label{fig:vector}
\end{figure}

\begin{table}
\begin{tabular}{crrdd}
& \multicolumn{2}{c}{$E_{c.m.}=0.432$~MeV} &
  \multicolumn{2}{c}{$E_{c.m.}=2$~MeV}  \\
& \multicolumn{1}{c}{Born} & \multicolumn{1}{c}{Full} &
  \multicolumn{1}{c}{Born} & \multicolumn{1}{c}{Full} \\
\hline
$^4S_{3/2}$ & -28.9 (-28.9)& -28.2 (-28.2)& -55.3(-55.3) & -63.1 (-63.1) \\
$^2P_{1/2}$ & -2.16 (-2.16)& -2.01 (-2.01)& -9.09(-9.10) & -7.36 (-7.37) \\
$^4P_{1/2}$ &  4.76 ( 4.76)&  5.30 ( 5.30)&  18.5(18.5)  &  22.1 (22.3)  \\
$\varepsilon_{1/2}^-$
            &  1.64 ( 1.65)&  2.45 ( 2.50)&  3.31(3.33) &  5.71  (5.83) \\
$^2P_{3/2}$ & -2.15 (-2.15)& -1.99 (-1.99)& -8.91(-8.93)& -7.14  (-7.15)\\
$^4P_{3/2}$ &  5.40 ( 5.39)&  6.16 ( 6.17)&  20.6(20.6) &  24.2  (24.2) \\
$\varepsilon_{3/2}^-$
            &-0.574(-0.573)&-0.861(-0.888)& -1.40(-1.40)& -2.20  (-2.23)\\
$^4P_{5/2}$ &  5.23 ( 5.23)&  5.78 ( 5.79)&  20.5(20.5) &  23.9  (24.1) \\
\end{tabular}
\caption{Some of the theoretical phase shifts and mixing parameters
(in degrees) calculated at $E_{c.m.}=0.432$ and 2~MeV for the AV18
potential; the values in parentheses correspond to the AV18+UR
potential. For both energies, the results for the ``optimized'' Born
approximation and  the full solution are reported in the columns
labeled ``Born'' and ``Full'', respectively.}
\label{tab:phase}
\end{table}

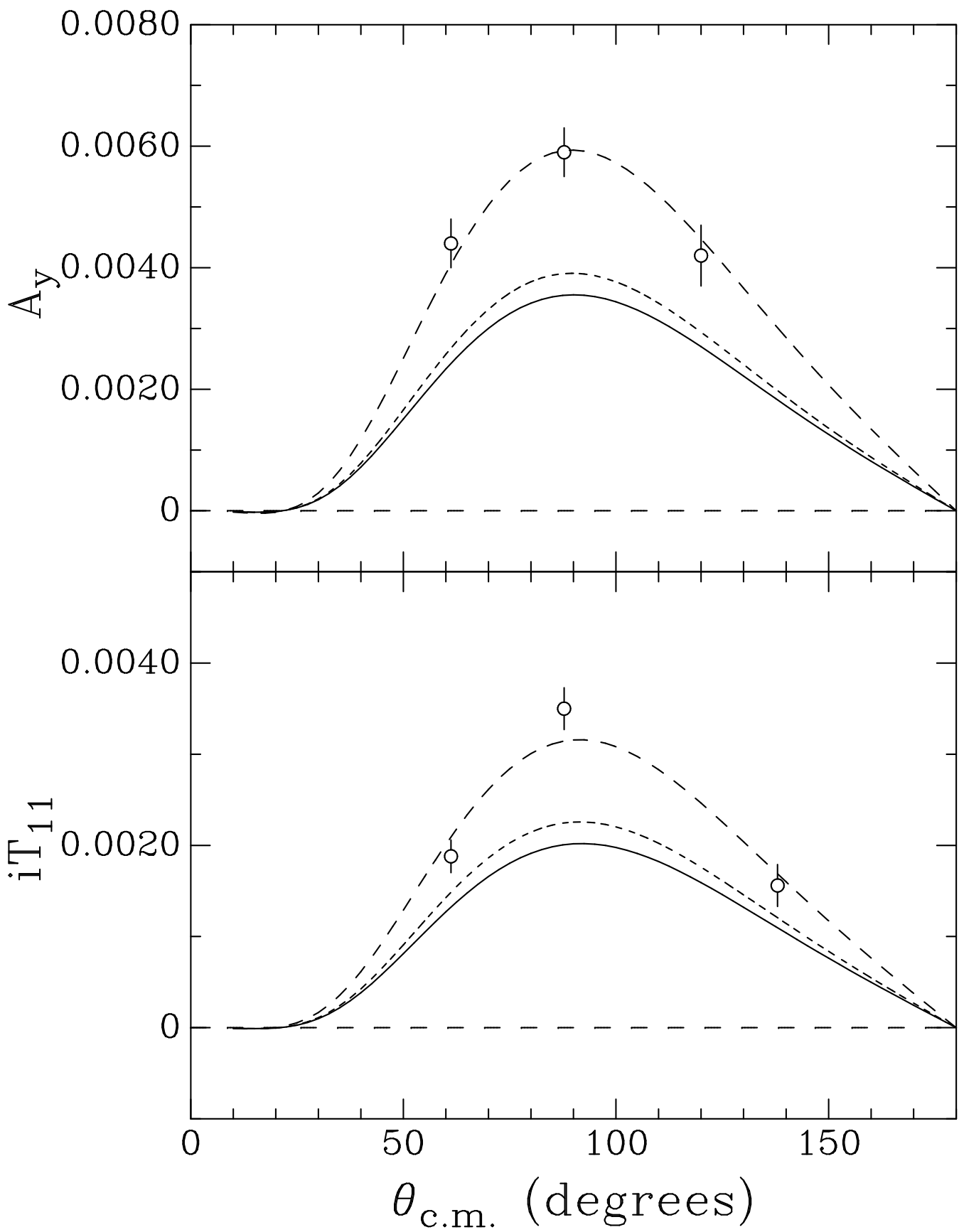
\end{document}